\documentclass{elsart}
\usepackage[dvips]{graphicx}
\usepackage[usenames]{color}

\def\diag{\mbox{\rm diag}}

\newtheorem{example}{Example}

\begin{document}
\begin{frontmatter}

\title{A Star Based Model for the Eigenvalue Power Law of Internet Graphs}

\author{Francesc Comellas\corauthref{fc}}
\corauth[fc]{Corresponding author. Tel. +34-93-413-7072, Fax
+34-93-413-7007.}
\ead{comellas@mat.upc.es}
\author{, Silvia Gago}
\address{
Departament de Matem\`atica Aplicada IV,\\
Universitat Polit\`ecnica de Catalunya,\\
08680 Castelldefels, Catalonia, Spain}

\begin{abstract}
Using  a simple deterministic model for the Internet graph
we  show that the  eigenvalue power law distribution for its adjacency matrix
is a direct consequence of the degree distribution and that the
graph must contain many star subgraphs. 
\end{abstract}

\date{16 Deember 2004}
\end{frontmatter}


\section{Introduction}
Recent research shows that most communications networks, 
like the world wide web, the Internet, telephone networks, 
transportation systems (including the power distribution network), 
and biological and social networks, belong to a class of networks 
known as small-world scale-free networks. 
These networks exhibit both  strong local clustering (nodes have many mutual neighbors) 
and a small diameter (maximum distance between any two nodes)~\cite{WaSt98}.
Another important common characteristic is that the number of links attached 
to the nodes usually obeys a power law distribution (is scale-free)
as was observed first in empirical studies  from the
Faloutsos' brothers \cite{FaFaFa99} and Bar\'abasi and Albert
\cite{BaAl99}. 
In \cite{FaFaFa99}, the authors present data that show that
the eigenvalue distribution of the adjacency matrix of the
Internet graph also follows a power law.
A qualitative proof that this power-law is a consequence from
the distribution of degrees of the network, which should
contain many star subgraphs of different sizes,
was presented by the authors in \cite{Co01}.
Mihail and Papadimitriou in~\cite{MiPa02} provided
an analytical proof by using the theory of random graphs.
Fabrikant, Koutsoupias and Papadimitriou~\cite{FaKoPa02} provide
an explanation of the Internet power laws based on  complex multicriterion optimization.

Many approaches to understanding small-world scale-free networks
are based on stochastic models and computer simulations, however
the use of deterministic models (although they do not allow
to capture the full complexity of a real life network) do help in
the understanding of their behavior and permit the 
direct determination of relevant parameters for the modeled networks.
In particular, it is possible to construct small-world deterministic graphs
with different degree distributions matching the distribution of 
real networks, see~\cite{CoOzPe00,BaRaVi01,CoSa02,CoFeRa04}.
We introduce in this paper a simple deterministic model, a toy model, to  show 
analytically, that the observed eigenvalue power law of the Internet may be a direct 
consequence of the degree distribution of a star based structure 
(whose power law can be justified considering, for example, 
preferential attachment~\cite{BaAl99} or duplication~\cite{ChLuDeGa03} models).

\section{A star graphs based deterministic model}
Restricting our study to the Internet graph (at the router level), 
our model considers this graph as a scale-free network made by 
the union of a number of star graphs with different orders
and connected through the root vertices by a
relatively small backbone graph $B$, see Figure 1.
\begin{figure}
\begin{center}
\includegraphics[width=8cm]{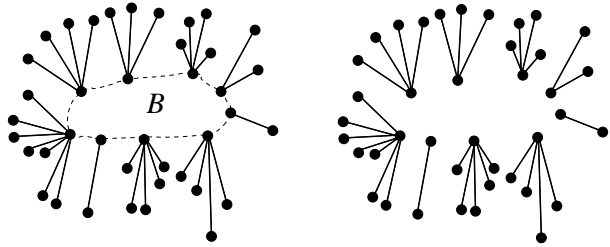}
\end{center}
\caption{ (a) A simple model for the Internet graph: many degree one nodes
are attached to a backbone graph $B$ to form graph $G$. 
(b) Deleting the backbone graph $B$ results in a non connected graph
constituted by the union of star graphs.}
\label{Fig1}
\end{figure}
This model is an approximation to the real Internet graph,
as described in appendix A of \cite{FaFaFa99}, if we consider a 
degree distribution matching the observed results:

Each $n_i+1$-star graph (with a root node and $n_i$ other nodes) will contribute
with a vertex of degree $n_i$ and $n_i$ vertices of degree $1$.
The eigenvalues of this star graph are $\pm\sqrt{n_i}$ and $0$.

If we consider a set of star graphs whose root vertices
have rank degrees distributed according
to a power law $\beta$, the graph union of these graphs will have
eigenvalues with ranks
distributed according to $\frac{\beta}{2}$. We note that, as we are 
considering a determinitic exact model, to relate the exponent of the 
discrete degree distribution to the 
$\beta$ exponent of a continuous degree distribution for a random scale-free graph, 
a cumulative distribution should be considered, 
$P_{cum}(k) \equiv \sum_{k^\prime \geq k} n(k^\prime)/N \sim k^{1-\beta}$,  where $k$ and $k^\prime$ are values of the discrete degree spectrum, $n(k^\prime)$ is the number of
vertices of degree $k^\prime$ and $N$ is the order of the graph.

It is possible now to give some insight on the eigenvalues
of the global graph $G$ obtained by joining the star graphs 
by using a backbone  graph $B$. Since the number 
of vertices of $B$ is small with 
respect to the total number of vertices of all the star graphs, the
values of the spectrum will be very close to the original
graph of stars and will therefore follow a power law.

This last result is related to the interlacing theorem~\cite{CvDoSa95}
which states that if $\Gamma$ is a graph with spectrum $\lambda_1 \geq \lambda_2 \geq \dots
\geq \lambda_p$,  $v_1$ is a vertex of $\Gamma$,  and 
$\mu_1 \geq \mu_2 \geq \dots \geq \mu_{p-1}$ is the spectrum of 
$\Gamma \setminus v_1$ (the graph resulting from $\Gamma$ after deleting
$v_1$ and its associated edges) then the spectrum of $\Gamma \setminus v_1$ is interlaced
with the spectrum of $\Gamma$, i.e.
$\lambda_1 \geq \mu_1 \geq  \lambda_2  \geq \mu_2 \geq  
\dots \geq \mu_{p-1} \geq \lambda_p$.

On the other hand, notice that this simple model ensures
that if the power-law exponent for the degrees is $\beta$,
the corresponding exponent for graph $G$ will be 
approximately   $\beta/{2}$, matching the
results shown in ~\cite{FaFaFa99}.

In the next section we compute analitically the eigenvalues of $G$ and
verify that they are close to  those of its star subgraphs.

\section{Analytical computation of the eigenvalues of $G$}
Consider $s$ star graphs $S_{1},S_{2},\dots,S_{s}$ with, respectively,
$n_{1}+1,n_{2}+1,\dots, n_{s}+1$ vertices.
The backbone graph $B$  joins the central
vertices of each star. We call $A$ the adjacency matrix of this graph $B$.
This is equivalent to start with a graph $B$ with $s$ vertices and
add to each vertex, respectively, $n_{1},\dots,n_{s}$ new vertices of degree 1.

To compute the spectrum of the global graph $G$ ($B$ plus the vertices
of degree one) we consider 
$\left( \begin{array}{cc}
        A & b^{T} \\
        b & O
        \end{array} \right)
\left( \begin{array}{c}
       x \\ y \end{array} \right) =
   \lambda   \left(\begin{array}{c}
        x \\ y \end{array} \right)$
which gives the system to solve 
\begin{equation}
 \Bigg\{
   \begin{array}{l}
    A x + b^{T} y = \lambda x \\
    b x  = \lambda y  \label{eq:sistema}
   \end{array}
\end{equation}
where $b^{T}=\diag(b_{1},\dots,b_{s})$ with $b_{i}=(1,\dots,1)$, $n_i$ ones,
i.e. $b^{T}$
is a $s \times (\sum_{i=1}^{s}n_{i})$ matrix and $O$
is the null matrix with dimensions $(\sum_{i=1}^{s}n_{i}) \times (\sum_{i=1}^{s}n_{i})$.

    If $\lambda \neq 0$, from (\ref{eq:sistema}) we obtain $A x +  \lambda^{-1} b^{T}b x  = \lambda x $
where $b^{T} b$ is a  diagonal matrix, whose elements are
$n_{1},\dots,n_{s}$. Thence $A x = (\lambda I - \lambda^{-1}\diag(n_{1},\dots,n_{s}))x$.
Introducing $\alpha_{i}=   \lambda/(\lambda^{2}-n_{i})$, this last equation
can be written
\begin{equation}
\diag(\alpha_{1},\dots,\alpha_{s}) A x=x.   \label{eq:tosolve}
\end{equation}

If the backbone graph $B$ is another star graph with $s$ nodes (rooted in one
of the star graphs), then
the matrix $A$ will be  $\left( \begin{array}{cc}
             0 & b^{T} \\
             b & O_{s-1}
    \end{array} \right)$
where $b^{T}=(1,\dots,1)$ and $O_{s-1}$
is a null matrix of dimensions $(s-1) \times (s-1)$. 
Thus, the eigenvalues of $G$  are the values of $\lambda$ 
wich verify  (\ref{eq:tosolve}) for this $A$, i.e.  values which allow
 the matrix $\diag(\alpha_{1},\dots,\alpha_{s}) A$ to have 1 as eigenvalue.

Solving $\det(\diag(\alpha_{1},\dots,\alpha_{s}) A-xI)=0$ we find
\begin{equation}
      (-x)^{s-2}\left(x^{2}-\alpha_{1}\sum_{i=2}^{s}\alpha_{i} \right)=0
          \label{eq:thealpha}
\end{equation}
and the eigenvalues are $0$ with multiplicity
$s-2$, and $\pm\sqrt{\alpha_{1}\sum_{i=2}^{s}\alpha_{i}}$ 
with multiplicity $1$. 
Hence,  the final equation to solve is $\alpha_{1}\sum_{i=2}^{s}\alpha_{i}=1.$
Its left part is a function of $\lambda$
$$\begin{array}{lll} 
f(\lambda) &=&  \alpha_{1}\sum_{i=2}^{s}\alpha_{i}=\sum_{i=2}^{s}\alpha_{1}\alpha_{i}=
     \sum_{i=2}^{s} \frac{\lambda}{\lambda^{2}-n_{1}}
    \frac{\lambda}{\lambda^{2}-n_{i}}\\
    &= &\sum_{i=2}^{s} \frac{n_{1}}{(n_{1}-n_{i})(\lambda^{2}-n_{1})}
     - \frac{n_{i}}{(n_{1}-n_{i})(\lambda^{2}-n_{i})}
\end{array}
$$
and it can be decomposed in a sum of simple fractions whose
asymptotes are given by  $\lambda_{i,\pm}=\pm
\sqrt{n_{i}}$. The points $\lambda'_{i}$ where this rational
function cuts the constant function $1$ are the solutions of
the equation, and they are therefore the eigenvalues of $G$ we are looking for,
see Figure~2.
\begin{figure}[htbp]
\begin{center}
\includegraphics[width=12cm]{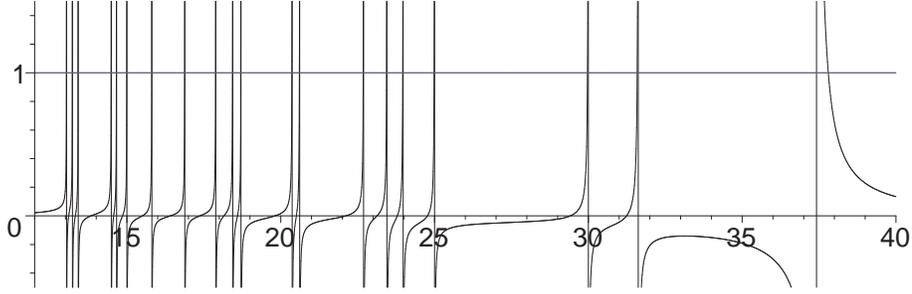}
\end{center}
\caption{Analytical determination of the eigenvalues of graph $G$, the graph obtained by joining several star graphs with a backbone graph.}
\label{Fig1}
\end{figure}

Each fraction $\frac{1}{\lambda^{2}-n_{i}}$ can be decomposed as 
    $\frac{1}{2\sqrt{n_{i}}(\lambda-\sqrt{n_{i}})}-
    \frac{1}{2\sqrt{n_{i}}(\lambda+\sqrt{n_{i}})}$.
If we consider the positive values to find the largest eigenvalues
and assuming $n_{1}\geq \dots \geq n_{s}$, for $2 \leq i \leq s $
approaching the value $\sqrt{n_{i}}$ from the right, the function $f(\lambda)$
is negative, whereas approaching it from the left it is positive.
Therefore it should cut the constant function $1$ in the left side, 
so $\lambda'_{i}<\sqrt{n_{i}}$, 
and  $\lambda'_{i}>\sqrt{n_{i+1}}$ as at $\sqrt{n_{i+1}}$ there is another 
asymptote and the function
is negative on the right side of $\sqrt{n_{i+1}}$. 
Therefore the  eigenvalues of $G$ verify:
$\sqrt{n_{i}}>\lambda'_{i}>\sqrt{n_{i+1}}$ and $\lambda'_{i}=
\sqrt{n_{i}}-\Delta$, where $\Delta<\sqrt{n_{i}}-\sqrt{n_{i+1}}$. 
At $\sqrt{n_{1}}$ the function is
positive at the right of this point and negative at the left.
Hence,  $\lambda'_{1}\geq \sqrt{n_{1}}$.
\newpage
Note that if the values for $n_i$ are consecutive integers,
the difference in the corresponding square roots will
be very small for large $n_i$ and the eigenvalues for the new graph
will be very near to those of the disconnected star graphs.

\begin{example} 
Consider the Oregon-Multi dataset of Internet at the AS level as in 
2001~\cite{SiFaFaFa03},  and use 
a star graph as backbone graph $B$ to interconnect the 
star graphs with $n_i$ equal to  1400, 1000, 900, 625, 575, 550, 515, 
425, 415, 350, 340, 320, 285, 250, 225, 215, 210, 180, 175 and 170.
The degree distribution (by construction) fits 
the degree distribution of the Internet graph as shown in~\cite{SiFaFaFa03}. 
To calculate the spectra of this graph we solve  the equation $f(\lambda)=1$ and
we obtain:  13.03, 13.22, 13.41, 14.48, 14.66, 14.99, 15.81, 16.87, 17.88, 18.43, 18.70, 
20.36, 20.61, 22.68, 23.44, 23.96, 24.98, 29.97, 31.59, 37.80.

In Figure 3 we plot the twenty highest eigenvalues 
in a log-log scale. The result shows a power law as in 
the real data  from ~\cite{SiFaFaFa03}. Observe 
that the slope of the regression model  is 
very similar to the observational data. 
\end{example}

\begin{figure}[h]
\begin{center}
\includegraphics[width=6cm]{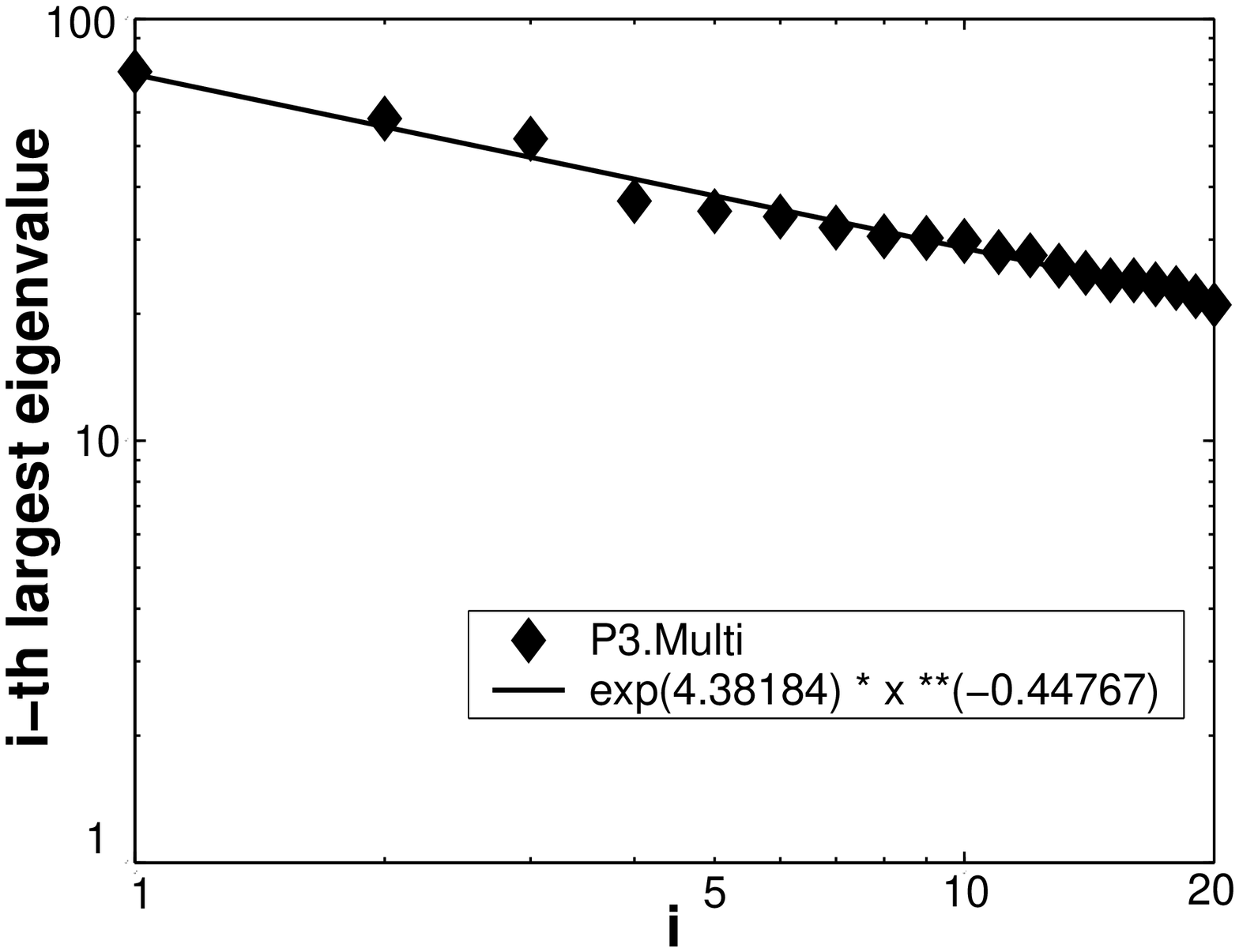}
\hskip 1cm
\includegraphics[width=6cm]{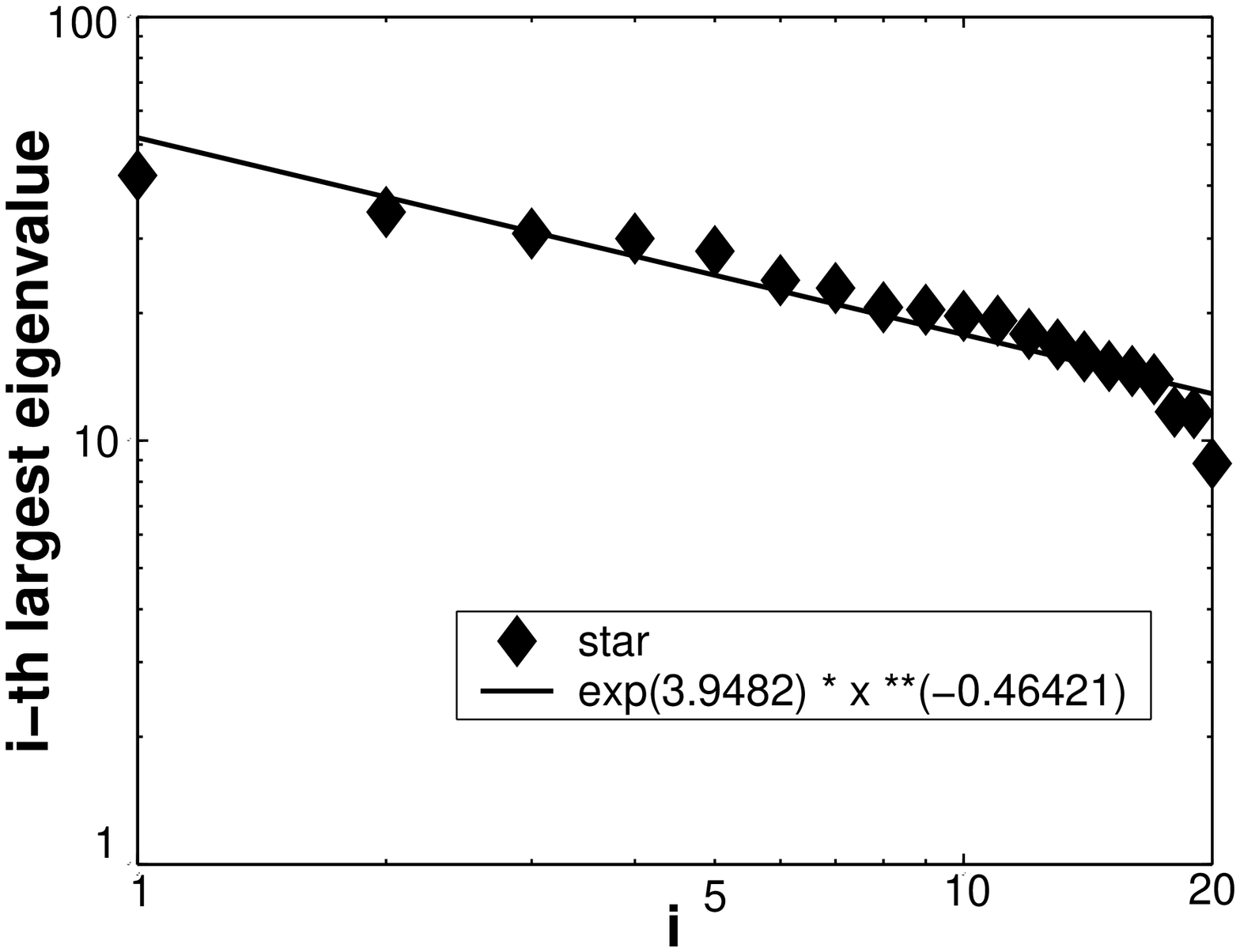}
\end{center}
\caption{Highest twenty eigenvalues for the Internet graph (Oregon-Multi), see~\cite{SiFaFaFa03}, and
the graph constructed in Example 1. In both cases they follow a power law.}
\label{Fig1}
\end{figure}

\section{Other models and conclusions}
The model considered here describes the degree and eigenvalue 
distributions of the Internet graph, however its clustering 
is zero as the 
resulting graph is a tree.
We have performed the same analysis connecting the root vertices
of the star graphs with a complete graph.  The results are 
comparable to those obtained above (in that case the clustering
of the global graph is different from zero but small).
We note that if the backbone graph $B$ joins complete graphs 
instead of star graphs this would lead
to an eigenvalue power law with an exponent close to the
exponent of the degree distribution (the eigenvalues of 
a complete graph with $n$ vertices are $n-1$ and $-1$).

The star-based model considered here constitutes a convenient tool to
study the Internet network since using deterministic techniques 
no simulation is needed and relevant network parameters 
can be directly calculated. Thus, the contrast of
real with synthesized networks is straightforward. 
On the other hand, our study shows that it is
a reasonable assumption to consider that the Internet graph
contains many star subgraphs, as this explains the observed
power-laws and relationship
between the degree and eigenvalue distributions.

\subsection*{Acknowledgment}
This research was supported by the Secretaria de Estado de Universidades 
e Investigaci\'on (Ministerio de Educaci\'on y Ciencia),  Spain, and the 
European Regional Development Fund (ERDF)  under project TIC2002-00155.


\end{document}